\documentclass[reprint,amsmath,amssymb,aps,onecolumn]{revtex4-2}
\usepackage{graphicx}
\usepackage[hidelinks]{hyperref}

\begin{document}

\title{Effect of finite {R}eynolds number on self-similar crossing statistics and fractal measurements in turbulence}

\author{Michael Heisel}
\email{heisel@ucla.edu}
\affiliation{Department of Atmospheric and Oceanic Sciences, University of California in Los Angeles, Los Angeles, USA}

\date{\today}

\begin{abstract}
Stochastic simulations are used to create synthetic one-dimensional telegraph approximation (TA) signals based on turbulent zero crossings, where the interval between crossings is governed by a power law probability distribution with exponent $\alpha$. The power law exponent is determined for statistics of simulated TA signals, namely the box-counting fractal dimension $D_1$, energy spectrum exponent $\beta_{TA}$, and an intermittency exponent $\mu_{TA}$. For the binary TA signal with no variability in amplitude, the parameters are related linearly as $D_1 = 2 - \beta_{TA} = 1 - \mu_{TA}$. The relations are unchanged if the crossing interval distribution has a finite power law region (i.e. inertial subrange) representing a flow with finite Reynolds number. However, the finite distribution yields statistics that are not truly scale-invariant, and distorts the linear relation between the statistic exponents and $\alpha$. The behavior is due to finite-size effects apparent from the survival function, or the complementary cumulative distribution, which for finite Reynolds number is only approximately self-similar and has an effective exponent differing from $\alpha$. An expression presented for the effective exponent recovers the expected relations between $\alpha$ and the TA statistics. The findings demonstrate how a finite Reynolds number can affect indicators of self-similarity, fractality, and intermittency observed from single-point measurements.
\end{abstract}

                              
\maketitle

\section{Introduction}
\label{sec1}

Turbulent fluid dynamics is one of relatively few fields where the existence of self-similarity (scale-invariance) is supported by both theory \cite{Kolmogorov1941} and extensive observation \cite{Pope2000}. The most well-known power law in turbulence describes self-similarity within the energy spectrum: for intermediate scales known collectively as the inertial subrange, the fluctuating energy decays as $E \sim f^{-\beta}$. Here $f$ is the frequency (or wavenumber) and $\beta$ is the spectral exponent. From a statistics perspective, mechanistic concepts underlying a power law include random walks \cite{Montroll1965,Newman2007}, fractal geometries \cite{Mandelbrot1982,Voss1985}, and self-organized criticality \cite{Bak1987,Bak1988,Jensen1998}.

Attempts to relate self-similarity in turbulence to concepts such as fractal geometries have provided both promising \cite[e.g.,][]{Mandelbrot1974,Sreenivasan1991} and conflicting \cite{Miller1991,Praskovsky1993,Catrakis1996a,Villermaux1999} evidence, where the latter studies observed a scale-dependent fractal dimension for isosurfaces and iso-crossings. The conflicting evidence may be explained by a combination of several challenges in isolating specific self-similar features in turbulence. For isosurfaces and iso-crossings near the mean value \citep{Miller1991,Praskovsky1993}, diffusive events are over-represented compared to level sets farther from the mean, leading to a relatively narrower inertial subrange of scales \cite[see, e.g.,][]{Iyer2020}. The same study \citep{Iyer2020} also demonstrated how scalar ramp-cliff patterns can influence the box-counting fractal dimension estimated from scalar concentration fields \citep{Miller1991,Catrakis1996a,Villermaux1999}.

An additional factor -- and the focus of the present study -- is the restriction of the self-similar behavior to a finite range of scales based on the Reynolds number $Re$. The finite inertial subrange is a form of truncated power law, where finite-size effects cause cumulative statistics to deviate from a true power law \cite{Pickering1995,Burroughs2001}. Previous works have briefly mentioned how finite-size effects can lead to an apparent scale-dependent fractal dimension \cite{Catrakis1996b,Catrakis2000}, but these effects have not otherwise been closely evaluated for power laws in turbulence. Specifically, it is not clear how the extent of the inertial subrange influences the relation between power law exponents of various statistics including the energy spectrum and fractal dimension.

One strategy to assess power law relations in turbulence is to reduce the measured fluctuating quantity to a one-dimensional crossing signal. The properties of isosurface geometries are ideally evaluated in three dimensions, but the reduction of the isosurface to its crossings of a one-dimensional transect is often an experimental necessity, particularly for point measurements in the high-Reynolds-number atmospheric surface layer. While the signal can be constructed based on crossings of any arbitrary level set value, the most common signal is defined using zero crossings of the velocity fluctuations \cite{Liepmann1949,Sreenivasan1983,Kailasnath1993}. The telegraph approximation (TA) signal \cite{Bershadskii2004,Cava2012} based on these zero crossings is 1 when the fluctuating velocity is positive, and is 0 when the velocity is negative. The interval between crossings is known as the interpulse period \cite{Sreenivasan1983,Cava2012} or persistence \cite{Perlekar2011,Chamecki2013,Chowdhuri2020}. The inertial subrange of the full velocity signal is similarly present in the zero-crossing TA signal. Among other statistics, the spectrum of the TA signal and the probability distribution of the interpulse period both exhibit self-similarity in the inertial subrange for sufficiently large $Re$ \cite[see, e.g.,][]{Bershadskii2004,Sreenivasan2006,Cava2012,Huang2021}.

Available measurements of simplified turbulent crossing signals present their own challenges for studying finite-size effects. First, a limited range of $\beta$ values is observed from turbulent measurements, which precludes empirical fits across the parameter space of $\beta$ and other exponent values. Second, even high-Reynolds-number flows have a narrow inertial subrange in the context of finite-size effects, as will be seen in later results. For instance, measurements in atmospheric flows typically have Taylor miscroscale Reynolds number $Re_\lambda \sim O(10^3)$, corresponding to no more than three decades (i.e. orders of magnitude) of self-similar inertial subrange \cite{Pope2000}.

To properly explore the full parameter space of Reynolds number and power law exponents, synthetic TA signals are constructed here using stochastic simulations based on idealized interpulse distributions. Power law statistics are computed across a range of interpulse exponent values, and the effective Reynolds number is also varied by truncating the power law region of the interpulse distributions. The statistics evaluated here are the fractal box-counting dimension, the energy spectrum, and an intermittency parameter. The simulations are analogous to a Monte Carlo analysis, except the goal is to identify the ensemble average of statistics rather than their uncertainty. The idealized simulations are purely stochastic and do not directly model any governing physics. This approach assumes the original signal is self-similar and identifies the consequence of a finite Reynolds number (truncated power law) on crossing statistics of the signal. While the analysis is discussed in the context of turbulent flows, the findings are generally applicable to any finite self-similar process.

The study is organized into the following sections: Sec. \ref{sec2} describes the stochastic simulations, Sec. \ref{sec3} presents results of the simulations, Sec. \ref{sec4} introduces a correction for finite Reynolds number, and Sec. \ref{sec5} summarizes the findings.

\section{Stochastic simulations}
\label{sec2}

The premise of the stochastic simulations is to create a synthetic TA signal $s(t)$ defined by a sequence of ``events'', where the interval $\tau$ between events is governed by a power law probability density function (PDF). In the context of a turbulent zero-crossing signal, each event represents the position $t$ (in space or time) where the fluctuating quantity $s$ crosses zero. The design of the simulations is detailed in the sections below for both ``unbounded'' and truncated interpulse power laws. The unbounded power law is used as a control case, and the truncated power laws approximate the interpulse distribution for three Reynolds numbers spanning the range $Re_\lambda \sim O(10^2-10^4)$.

\subsection{Unbounded power law}
\label{sec2_A}

\begin{figure}
\centering
\includegraphics{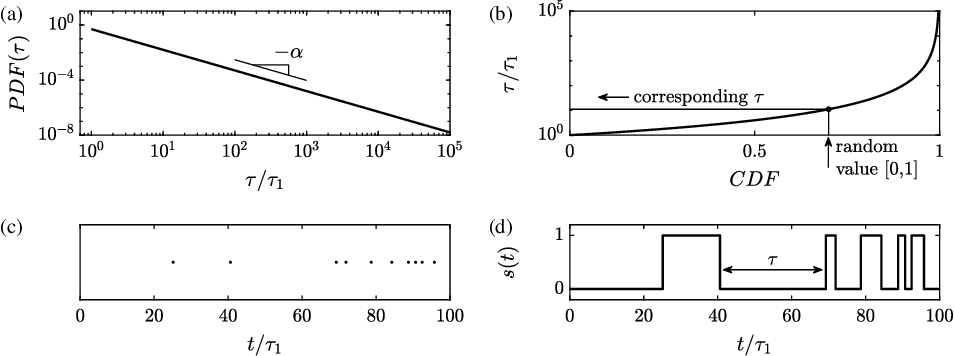}
\caption{Example synthetic telegraph approximation (TA) signal constructed from a power law probability distribution $PDF \sim \tau^{-\alpha}$. (a) Probability distribution for the interval $\tau$ between events in a signal following Eq. \eqref{eq1}. (b) Determination of a single $\tau$ value using inverse transform sampling of the cumulative distribution $CDF$ and Eq. \eqref{eq3}. (c) Position $t$ of events based on ten $\tau$ values. (d) TA signal $s(t)$ whose value 0 or 1 changes at each event.}
\label{fig1}
\end{figure}

For the unbounded control case, the TA interpulse distribution is modeled as a power law that exceeds the extent of the simulation domain. The power law PDF is defined as

\begin{equation}
PDF(\tau) = \frac{1}{1-\left(\tau_2 / \tau_1 \right) ^{1-\alpha} } \frac{\alpha-1}{\tau_1} \left( \frac{\tau}{\tau_1} \right)^{-\alpha},
\label{eq1}
\end{equation}

\noindent where $\tau_1$ and $\tau_2$ are the minimum and maximum values of the distribution, respectively, and $\alpha$ is the distribution exponent. The power law in Eq. \eqref{eq1} is equivalent to a Pareto distribution with exponent $\alpha-1$. The integral of the PDF is equal to unity -- as required by the PDF definition -- only for $\alpha >$ 1 and if a minimum value is imposed. The integral is infinite and the PDF is not well-defined for $\alpha \le$ 1. The minimum $\tau_1=$ 1 is used here for simplicity. While a maximum value $\tau_2$ is typically $\infty$ for an unbounded power law, $\tau_2=10^{300}$ is employed to avoid infinite values in the simulations. This value is close to the largest definable number in double-precision floating-point format, and yields 300 decades of power law. An example PDF is shown in Fig. \ref{fig1}(a).

Inverse transform sampling is used to select values of $\tau$ from the distribution. In this approach, the cumulative distribution function

\begin{equation}
CDF(\tau) = \frac{1}{1-\left(\tau_2 / \tau_1 \right) ^{1-\alpha} } \left[ 1 - \left( \frac{\tau}{\tau_1} \right)^{1-\alpha} \right]
\label{eq2}
\end{equation}

\noindent is inverted to define $\tau$ as a function of $CDF$:

\begin{equation}
\tau(CDF) = \tau_1 \left[ 1 - \left( 1 - \left( \frac{\tau_2}{\tau_1} \right) ^{1-\alpha} \right) CDF \right] ^\frac{1}{1-\alpha}.
\label{eq3}
\end{equation}

The $CDF$ value is simulated by selecting a random value between 0 and 1, and the corresponding interval $\tau$ is determined from Eq. \eqref{eq3} as shown in Fig. \ref{fig1}(b). A small sample of events is shown in Fig. \ref{fig1}(c), where each event is separated by simulated intervals $\tau$.

The position $t$ of each event is used to build a synthetic TA signal $s(t)$. Following the TA definition given in the introduction, the $s(t)$ value alternates between 0 and 1 at each event as seen in Fig. \ref{fig1}(d). The signal is constructed on a discrete domain between 0 and $t_{max}=10^6$. The resolution between points on the domain is $\tau_1 =$ 1, thus allowing for six decades of statistics in the signal. While $PDF(\tau)$ represents 300 decades of power law, only six decades can be observed in $s(t)$ due to the limited domain. However, based on later results, the effect of truncation by the domain size is negligible for this case such that it is considered unbounded for the purpose of the study.

Later results present simulations for one hundred $\alpha$ values between 1.02 and 3. The value $\alpha=1$ is excluded because the PDF is not well-defined as discussed above. The approximate exponent for a turbulent TA signal is $\alpha \approx 1.5$ \cite{Bershadskii2004}. For a given $\alpha$, intervals $\tau$ are simulated until $t_{max}$ is exceeded to ensure the signal is fully populated. Statistics are thereafter calculated using $s(t)$, and the process is repeated until the ensemble average statistics are converged. The number of signals contributing to each statistic varies between $10^2$ for large $\alpha$ and $10^4$ for small $\alpha$. The latter returns sparse signals, which require a larger number of realizations to converge statistics.

\subsection{Truncated power law}
\label{sec2_B}

\begin{figure}
\centering
\includegraphics{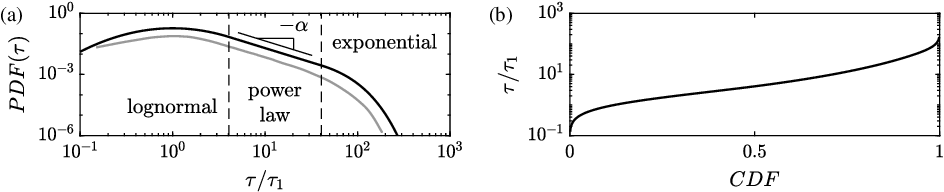}
\caption{Power law distribution whose extent is truncated by lognormal and exponential curves. The example exhibits one decade of power law, corresponding to Reynolds number $Re_\lambda \sim 10^2$. (a) Probability distribution for $\tau$ following Eq. \eqref{eq4}. The gray line is an example interpulse distribution from boundary layer turbulence measurements \cite{Heisel2020}, shifted for visual comparison. (b) Inverse transform sampling of $CDF$ following Eq. \eqref{eq7}.}
\label{fig2}
\end{figure}

For a turbulent TA signal, the interpulse PDF is only a power law for $\tau$ values within the inertial subrange of scales \citep{Bershadskii2004,Cava2012}. Smaller interpulse periods are well approximated by a lognormal distribution \cite{Badri1977,Bershadskii2006}, and larger $\tau$ values follow an exponential cutoff in boundary layer flows \cite{Sreenivasan1983,Cava2012,Chamecki2013}. The same PDF shape -- a blend of lognormal, power law, and exponential distributions -- is used here to simulate the effect of a truncated power law region. The case represents a weak power law because a power law expression does not describe the full range of values in the distribution \cite{Percline2005}.

Same as for the unbounded power law, inverse transform sampling is used to simulate $s(t)$ based on a random selection of values for $\tau$. For simplicity, the lognormal portion of the distribution is defined using the parameters $\mu_* =$ 1 and $\sigma_*^2 =$ 1, which respectively correspond to the mean and variance of log($\tau$). From these parameters, the mode of the lognormal curve is fixed at $\tau_1=$ 1. The resulting PDF for $\tau$ is given by the piecewise function

\begin{equation}
PDF(\tau) = 
\begin{cases}
	\frac{C_1}{\tau} e^\frac{-\left( \log(\tau) - 1 \right) ^2}{2}	& \tau \le e^\alpha 		\\
	C_2 \tau^{-\alpha}											& e^\alpha < \tau \le b	\\
	C_3 e^{-\lambda \tau}										& b < \tau \le \tau_2 	\\
\end{cases}
\label{eq4}
\end{equation}

\noindent The transition from the lognormal to the power law curve occurs at $e^\alpha$. This point corresponds to $d \log(PDF) / d \log(\tau) = \alpha$ along the lognormal curve, ensuring a smooth transition to the power law. The transition to the exponential cutoff is imposed at $b=10^x e^\alpha$, where $x$ is the desired number of power law decades. The exponential parameter $\lambda=\alpha / b$ enforces a smooth transition to the exponential cutoff, i.e.  $d \log(PDF) / d \log(\tau) = \alpha$ at $\tau=b$. The factors are defined as

\begin{eqnarray}
C_1 &=& \left[ C_4 + \frac{e^{\frac{\alpha^2-1}{\alpha}}}{1-\alpha} C_5 - \frac{e^{\frac{1}{2} \left(\alpha^2-1\right)+\alpha}}{\alpha b^{\alpha-1}} \left( e^{-\lambda \tau_2} - e^{-\alpha} \right) \right]^{-1}	\nonumber	\\
C_2 &=& C_1 e^{\frac{\alpha^2-1}{2}}	\nonumber	\\
C_3 &=& C_2 \left( \frac{e}{b} \right)^\alpha	\nonumber	\\
C_4 &=& \sqrt{\frac{\pi}{2}} \left( \mathrm{erf}\left( \frac{\alpha-1}{\sqrt{2}} \right) + 1 \right)	\nonumber	\\
C_5 &=& b^{1-\alpha} - e^{\alpha \left(1-\alpha\right)}.
\label{eq5}
\end{eqnarray}

\noindent The constants $C_2$ and $C_3$ are defined relative to $C_1$ to ensure the amplitude of $PDF(\tau)$ is matched at the transition points, and $C_1$ is defined to achieve $\int_{0}^{\tau_2} PDF(\tau) =$ 1. The cumulative distribution corresponding to Eq. \eqref{eq4} is

\begin{equation}
CDF(\tau) = 
\begin{cases}
	C_1 \sqrt{\frac{\pi}{2}}  \left( \mathrm{erf} \left( \frac{ \log(\tau) - 1}{\sqrt{2}} \right) +1 \right) & \tau \le e^\alpha 		\\
	C_1 C_4 + \frac{C_2}{1-\alpha} \left( \tau^{1-\alpha}-e^{\alpha (1-\alpha)} \right)		& e^\alpha < \tau \le b	\\
	C_1 C_4 + \frac{C_2 C_5}{1-\alpha} + \frac{C_3}{\lambda} \left( e^{-\lambda \tau} - e^{-\alpha} \right)	& b < \tau \le \tau_2 	\\
\end{cases}
\label{eq6}
\end{equation}

\noindent Finally, the inversion of Eq. \eqref{eq6} yields the transform equation used to simulate the finite power law:

\begin{equation}
\tau = 
\begin{cases}
	\exp \left[ \sqrt{2} \, \mathrm{erfinv}\left( \sqrt{\frac{\pi}{2}} \frac{CDF}{C_1} - 1 \right) + 1 \right]	&	CDF \le C_1 C_4	\\
	\left[ \frac{1-\alpha}{C_2} \left( CDF - C_1 C_4 \right) + e^{\alpha \left( 1-\alpha \right)} \right] ^{\frac{1}{1-\alpha}}	&	C_1 C_4 < CDF \le C_1 C_4 + \frac{C_2 C_5}{1-\alpha}	\\
	- \frac{1}{\lambda} \log \left[ \frac{\lambda}{C_3} \left( C_1 C_4 + \frac{C_2 C_5}{1-\alpha} - CDF \right) +  e^{-\alpha} \right]	&	CDF > C_1 C_4 + \frac{C_2 C_5}{1-\alpha}.	\\
\end{cases}
\label{eq7}
\end{equation}

\noindent In Eqs. \eqref{eq6} and \eqref{eq7}, erf($x$) and erfinv($x$) refer to the error function and its inverse, respectively, and the two notations for the exponential function $e^x$ and $\exp(x)$ are used interchangeably for readability.

Figure \ref{fig2} shows a synthetic truncated power law. An example turbulent zero-crossing (interpulse) PDF is included for reference. The interpulse is estimated from hotwire anemometry measurements of boundary layer turbulence \cite{Heisel2020}, and exhibits a similar shape to the simulated PDF with one decade of power law. Using the inverse transform of $CDF$ in Fig. \ref{fig2}(b) and Eq. \eqref{eq7}, the truncated signals are simulated on the same domain and for the same range of $\alpha$ values as the unbounded case.

While the synthetic and experimental distributions in Fig. \ref{fig2} appear to be qualitatively similar, the idealized distribution is not designed to reproduce aspects of a zero-crossing signal that are outside the scope of the work. For instance, the parameters for the lognormal region do not produce the correct scaling for the mean value of $\tau$ \cite{Sreenivasan1983}, and the direct transitions between the different scaling regions may not accurately reflect experimental observations.

The breadth of the simulated power law region can be related directly to the Reynolds number. Assuming the inertial subrange spans $O(10\eta)$ to $O(L)$ \cite{Saddoughi1994}, where $\eta$ and $L$ are the Kolmogorov microscale and the integral scale, respectively, the extent of the inertial subrange is $O(0.01 Re_\lambda^{3/2})$ \cite{Pope2000}. The number of decades in the inertial subrange is therefore $\tfrac{3}{2} \log_{10}(Re_\lambda)-2$. Results for distributions with one ($Re_\lambda \sim 10^2$), three ($Re_\lambda \sim 10^3$), and five ($Re_\lambda \sim 10^4$) decades of power law are presented herein. The six decades of resolution in the simulated domain yield $Re_\lambda \gtrsim 10^5$ for the unbounded case.

\section{Results}
\label{sec3}

\subsection{Fractal Dimension}

\begin{figure}
\centering
\includegraphics{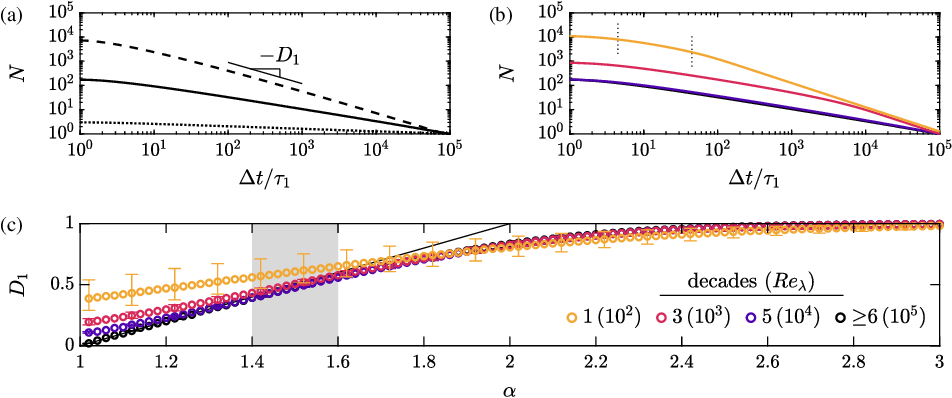}
\caption{Statistics for the fractal dimension $D_1$ estimated via box counting on simulated TA signals. (a) Box count $N \sim \Delta t^{-D_1}$ for the unbounded case and $\alpha = 1.1$ (dotted line), $\alpha = 1.5$ (solid), and $\alpha = 2$ (dashed). (b) Box count for $\alpha=1.5$ and varying Reynolds number, where the vertical lines delineate the power law region for the $Re_\lambda \sim 10^2$ case. (c) Fractal dimension $D_1$ as a function of $\alpha$ with $D_1=\alpha-1$ (line) for reference. The shaded region in (c) corresponds to the typical value $\alpha \approx 1.5$ for a turbulent TA signal. In this and later figures, the legend indicates the number of power law decades in $PDF(\tau)$ and the order of the equivalent Reynolds number $Re_\lambda$ based on the Taylor microscale.}
\label{fig3}
\end{figure}

The fractal dimension of the simulated signal is estimated here applying the box counting method \cite[see, e.g.,][]{Russell1980,Voss1985,Theiler1990} to the event positions featured in Fig. \ref{fig1}(c). In the box-counting approach, the domain is discretized into segments of size $\Delta t$ and the number of segments $N$ containing at least one event is counted. If the resulting dependency $N(\Delta t)$ follows a power law

\begin{equation}
N(\Delta t) \sim \Delta t^{-D_1},
\label{eq8}
\end{equation}

\noindent the signal is considered statistically self-similar. Whether the signal is also considered to be a fractal object depends on the definition, as fractality is sometimes reserved for geometric shapes. The subscript 1 is adopted for the fractal dimension $D_1$ in Eq. \eqref{eq8} because the estimate is made on a one-dimensional signal. The value for $D_1$ is bounded between 0 and 1. These limits correspond to a signal with $\tau > \Delta t$ (for $D_1=0$) or $\tau < \Delta t$ (for $D_1=1$) for all intervals $\tau$ across the tested range of $\Delta t$.

Example box-counting results are shown in Fig. \ref{fig3}(a,b) for a range of $\alpha$ and power law truncation. The expected power law in Eq. \eqref{eq8} is approximately observed for the unbounded distribution. However, in Fig. \ref{fig3}(b) $N(\Delta t)$ becomes increasingly dissimilar from a power law as the self-similar region in $PDF(\tau)$ is increasingly truncated. This trend is consistent with similar Monte Carlo simulations that showed $D_1(\Delta t)$ to vary with $\Delta t$ for any truncated power law due to finite-size effects \cite{Catrakis2000}. As a result, a constant fractal dimension can only be achieved in an approximate sense for finite Reynolds number flows. The absence of a true power law in Fig. \ref{fig3}(b) is further discussed in Sec. \ref{sec4}.

Despite the departure from a power law, the dimension $D_1$ is estimated by fitting Eq. \eqref{eq8} to the Fig. \ref{fig3}(a,b) curves assuming $D_1$ is constant. The fit is performed within the range of $\Delta t$ corresponding to the self-similar region in $PDF(\tau)$. The fitted values for $D_1$ across the tested range of $\alpha$ are shown in Fig. \ref{fig3}(c). The error bars correspond to the change in $D_1$ when the region where the power law is fitted is shifted by a factor of two in either direction. The error bars therefore increase as the dependence $D_1(\Delta t)$ increases. The large error bars corresponding to small $\alpha$ in Fig. \ref{fig3}(c) reflect the dissimilarity from a true power law observed in Fig. \ref{fig3}(b). The same method is used to calculate the error bars in later figures.

For the unbounded PDF case in Fig. \ref{fig3}(c), $D_1(\alpha)$ follows a linear trend $D_1=\alpha-1$ up to approximately $\alpha \approx$ 1.5, where the linear relation matches previous predictions \cite{Catrakis1996b,Catrakis2000}. Above $\alpha \approx$ 1.5, $D_1$ asymptotically approaches 1. The same asymptotic behavior is observed for the truncated distributions representing finite Reynolds number. However, the results for small $\alpha$ depart from the linear relation as the equivalent Reynolds number decreases. The reason for the departure is related to the trends in Fig. \ref{fig3}(b) and is discussed in Sec. \ref{sec4}.

\subsection{Energy spectrum}

The energy spectrum is defined as $E_{TA}(f) \sim \lvert \hat{s}(f) \rvert ^2$, where $\hat{s}(f)$ is the Fourier transform of $s(t)$ in frequency or wavenumber space. Prior to computing the transform, $s(t)$ is multiplied by a Hamming window filter whose length matches the domain size $t_{max}$. The signal is also zero-padded. The window filter and zero-padding mitigate aliasing in $\hat{s}(f)$.

\begin{figure}
\centering
\includegraphics{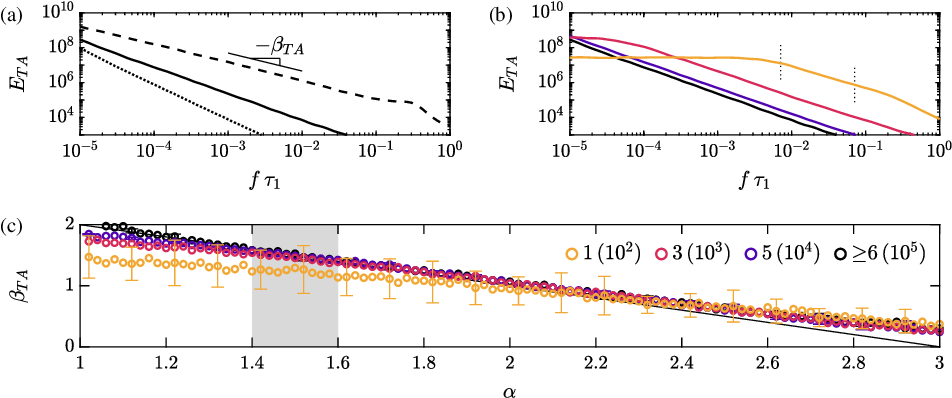}
\caption{Statistics for the energy spectrum power law exponent $\beta_{TA}$ of simulated TA signals. (a) Spectrum $E_{TA}$ for the unbounded case and $\alpha = 1.1$ (dotted line), $\alpha = 1.5$ (solid), and $\alpha = 2$ (dashed). (b) Spectrum for $\alpha=1.5$ and varying Reynolds number, where the vertical lines delineate the power law region for the $Re_\lambda \sim 10^2$ case. (c) Exponent $\beta_{TA}$ as a function of $\alpha$ with $\beta_{TA}=3-\alpha$ (line) for reference. The shaded region in (c) corresponds to the typical value $\alpha \approx 1.5$ for a turbulent TA signal.}
\label{fig4}
\end{figure}

Using the same format as Fig. \ref{fig3}, the resulting energy spectra are shown in Fig. \ref{fig4}. Spectra for the truncated distributions in Fig. \ref{fig4}(b) exhibit stronger self-similarity than the box counts in Fig. \ref{fig3}(b). The Fourier transform efficiently isolates local (in scale) contributions to the variance. In contrast, the box-counting measures a cumulative effect capturing all intervals smaller than the given $\Delta t$. For the cumulative statistics, the non-power-law behavior is spread across scales to the expected self-similar region.

The spectral exponent $\beta_{TA}$ is estimated by fitting the power law $E_{TA} \sim f^{-\beta}$ to each individual spectrum. The ``$TA$'' subscript is adopted because $\beta_{TA} \approx 4/3$ observed for turbulent flows differs from the value $\beta \approx 5/3$ for the full signal \cite{Sreenivasan2006}. The dependency of $\beta_{TA}$ on $\alpha$ is shown in Fig. \ref{fig4}(c). The apparent ``roughness'' of the curves is attributed to the shape of $s(t)$. Artificial oscillations appear in the energy spectrum when the sinusoidal basis functions of the Fourier transform are used to decompose the discontinuous signal. These oscillations may propagate to $\beta_{TA}$ as the fitted power law region varies with $\alpha$.

As before, the error bars are largest for the lowest equivalent Reynolds number due to the curvature of $E_{TA}(f)$ immediately adjacent to the expected self-similar region. The trend for small $\alpha$ observed in Fig. \ref{fig3}(c) is similarly present for the spectrum exponent. The unbounded case follows a linear relation $\beta_{TA}=3-\alpha$ in Fig. \ref{fig4}(c) up to $\alpha \approx$ 2. This relation is applicable to a superposition of Poisson processes \cite{Jensen1998}, and appears similarly applicable to the self-similar process simulated here. For $\alpha >$ 2, the results slowly deviate from the linear relation and there is agreement across cases. It is assumed that $\beta_{TA}$ asymptotically approaches 0 as $\alpha$ increases, but this trend cannot be confirmed due to the limited tested range of $\alpha$.

\subsection{Intermittency}

\begin{figure}
\centering
\includegraphics{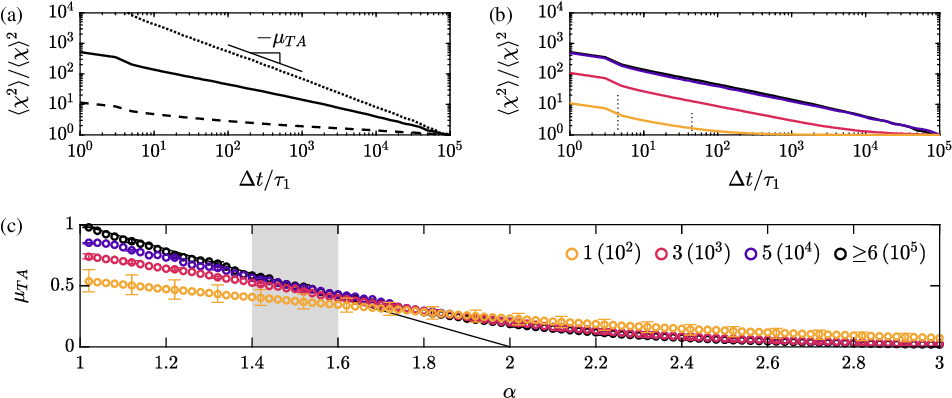}
\caption{Statistics for the intermittency exponent $\mu_{TA}$ of simulated TA signals. (a) Intermittency parameter $\langle \chi ^2 \rangle / \langle \chi \rangle ^2$ for the unbounded case and $\alpha = 1.1$ (dotted line), $\alpha = 1.5$ (solid), and $\alpha = 2$ (dashed). (b) Intermittency for $\alpha=1.5$ and varying Reynolds number, where the vertical lines delineate the power law region for the $Re_\lambda \sim 10^2$ case. (c) Exponent $\mu_{TA}$ as a function of $\alpha$ with $\mu = 2-\alpha$ (line) for reference. The shaded region in (c) corresponds to the typical value $\alpha \approx 1.5$ for a turbulent TA signal.}
\label{fig5}
\end{figure}

The intermittency is another quantitative measure of variability in the distribution of events. Intermittency can be parameterized using Obukhov's local moving average \cite{Obukhov1962}

\begin{equation}
\chi(t,\Delta t) = \frac{1}{\Delta t} \int_t^{t+\Delta t} \bigg| \frac{ds^2}{dt} \bigg| dt.
\label{eq9}
\end{equation}

\noindent Given the amplitude of $s(t)$ is invariable, the integral corresponds to the number of events occurring within ``windows'' of size $\Delta t$. The intermittency is quantified using the scaling \cite{Monin1975,Sreenivasan1997,Bershadskii2004}

\begin{equation}
\frac{ \langle \chi ^2 \rangle}{ \langle \chi \rangle ^2} \sim \Delta t^{-\mu_{TA}},
\label{eq10}
\end{equation}

\noindent where angled brackets $\langle \cdot \rangle$ indicate an ensemble average across $t$. Equation \eqref{eq10} is defined here using the second-order moment, but the same principle can be applied to higher-order moments. The exponent $\mu_{TA}$ represents how the variability in number of events across windows changes as the window size is increased.

The intermittency parameter statistics are shown in Fig. \ref{fig5}. Results for the truncated distributions in Fig. \ref{fig5}(b) exhibit a lack of self-similarity. Same as for the box counting methodology, Eq. \eqref{eq9} accounts for all intervals smaller than $\Delta t$, resulting in a cumulative metric.

Values for $\mu_{TA}$, fitted using Eq. \eqref{eq10}, are plotted as a function of $\alpha$ in Fig. \ref{fig5}(c). The curves follow the same trends as $D_1$ and $\beta_{TA}$. A linear relation $\mu_{TA} = 2 - \alpha$ is observed for the unbounded distribution and small $\alpha$ values. Truncating the power law distribution to represent a finite Reynolds number leads to a departure from the linear relation. For larger $\alpha$, all cases deviate from $\mu_{TA} = 2 - \alpha$ as $\mu$ asymptotically approaches zero.

\subsection{Exponent relations}

\begin{figure}
\centering
\includegraphics{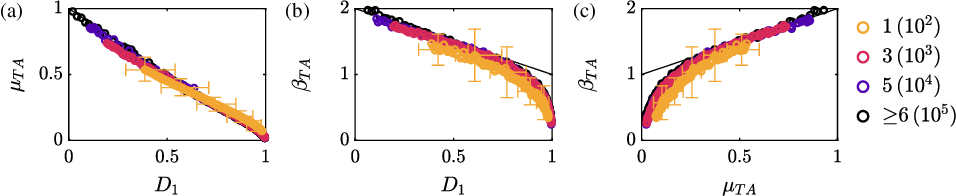}
\caption{Relations between power law exponents of the presented statistics for simulated TA signals. (a) Fractal dimension $D_1$ and intermittency exponent $\mu_{TA}$, compared to $\mu_{TA}=D_1-1$ (line). (b) Dimension $D_1$ and spectrum exponent $\beta_{TA}$, compared to $\beta_{TA}=2-D_1$ (line). (c) Exponents $\mu_{TA}$ and $\beta_{TA}$, compared to $\beta_{TA}=\mu+1$ (line).}
\label{fig6}
\end{figure}

The relations between power law exponents $D_1$, $\beta_{TA}$ and $\mu_{TA}$ are plotted in Fig. \ref{fig6} for the tested range of $\alpha$. The lines correspond to the linear trends in panel (c) of Figs. \ref{fig3}, \ref{fig4}, and \ref{fig5}. For visual clarity, the error bars from previous figures are only reproduced in Fig. \ref{fig6} for the truncated case with one decade of self-similarity.

A robust inverse relation between $\mu_{TA}$ and $D_1$ is observed in Fig. \ref{fig6}(a). The results are in close agreement with the prediction $\mu_{TA} = 1-D_1$ \cite{Frisch1978,Sreenivasan1997}, which can be derived via the correlation dimension \cite{Hentschel1983}. The linear trend is invariant to the Reynolds number (i.e. the truncation of the power law PDF), and the primary difference across cases is the observed range in $\mu_{TA}$ and $D_1$ values.

Linear trends $\beta_{TA}=2-D_1$ and $\beta_{TA}=\mu_{TA}+1$ also exist for the spectral exponent in Fig. \ref{fig6}(b,c). However, the linearity is limited to $\beta_{TA} \gtrapprox 1.2$. The behavior for smaller $\beta_{TA}$ is attributed to the slower rate at which $\beta_{TA}$ asymptotically approaches zero, compared to the corresponding rates for $D_1$ and $\mu_{TA}$. 

The results for $PDF(\tau)$ with $Re_\lambda \sim 10^2$ are visibly offset from the other cases in Fig. \ref{fig6}(b,c). The difference in $\beta_{TA}$ is approximately 0.2, which is within the extent of the error bars. The difference may therefore be due to the lack of self-similarity in the statistics and the precise range chosen to fit the power law exponents. The result highlights the challenge in recovering the expected relations when the Reynolds number yields a narrow inertial subrange and the cumulative statistics ($D_1$, $\mu_{TA}$) lose the signature of self-similarity.

Aside from the offset, the relations in Fig. \ref{fig6} do not depend on the Reynolds number and the bounds of the power law PDF. Direct linear relations can be expected between power law statistics, even if the governing distribution $PDF(\tau)$ is self-similar across a finite range of values. The effect of Reynolds number on the power law exponent relations is therefore limited to the altered connection between the statistics and the underlying probability exponent $\alpha$.

\section{Effect of finite Reynolds number}
\label{sec4}

The probability distribution definitions in Eqs. \eqref{eq1} and \eqref{eq2} impose a finite maximum value $\tau_2$. As a result, intervals above $\tau_2$ are under-sampled relative to an infinite power law distribution \cite{Pickering1995,Burroughs2001}. Values within the power law region may be under- or over-sampled, depending on the shape of the cutoff regions bounding the power law.

The consequence of the sampling discrepancy is apparent in the survival function $1-CDF(\tau)$, also known as the complementary cumulative distribution. For the power law in Eq. \eqref{eq2}, the survival function can be expressed as

\begin{equation}
1-CDF(\tau) = \frac{ \left( \tau / \tau_1 \right) ^{1-\alpha} - \left( \tau_2 / \tau_1 \right) ^{1-\alpha} }{ 1 - \left( \tau_2 / \tau_1 \right) ^{1-\alpha} }.
\label{eq11}
\end{equation}

\noindent Equation \eqref{eq11} is only a power law for infinite Reynolds number with $\tau_2/\tau_1 \rightarrow \infty$, and otherwise has a finite additive constant distorting the probability that the next interval exceeds a given value of $\tau$. The distortion has led to the use of more generalized distributions such as the Zipf--Mandelbrot law \cite{Mandelbrot1965} to describe cumulative statistics of truncated power laws.

\begin{figure}
\centering
\includegraphics{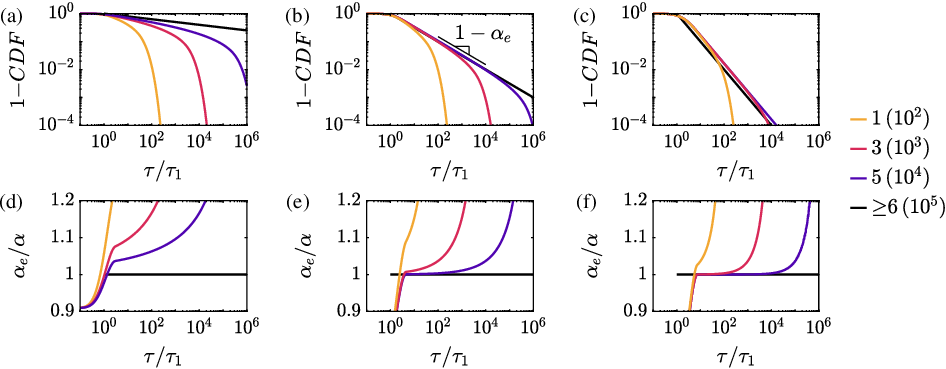}
\caption{Effective exponent $\alpha_e$ resulting from the distortion of the survival function by the truncated power law, i.e. finite Reynolds number. Rows correspond to the survival function $1-CDF(\tau)$ (a,b,c) and associated exponent $\alpha_e$ (d,e,f) following Eq. \eqref{eq:effective}. Columns correspond to values of $\alpha$: 1.1 (a,d); 1.5 (b,e); and 2 (c,f).}
\label{fig7}
\end{figure}

The survival functions for three values of $\alpha$ are shown in Fig. \ref{fig7}(a,b,c). The trends are consistent with the previously observed dependencies on $\alpha$: for finite Reynolds number with a truncated distribution the survival function only approximates a power law, and the departure from a power law is largest for small $\alpha$ values.

To quantify the deviation from a true power law, the effective exponent $\alpha_e$ can be calculated from the slope of the curves in Fig. \ref{fig7}(a,b,c). Mathematically, the exponent is

\begin{equation}
1-\alpha_e = \frac{d}{d \log(\tau)} \left[ \log \left( 1-CDF \right) \right].
\label{eq12}
\end{equation}

\noindent The chain rule can be used to simplify the derivative operation as $d/d\log(\tau) = \tau d/d\tau$. Using Eq. \eqref{eq11} to compute the derivative of $\log(1-CDF)$, the effective exponent can be expressed as

\begin{equation}
\alpha_e(\tau) = 1 + \frac{ \tau PDF(\tau) }{1-CDF(\tau)}.
\label{eq13}
\end{equation}

Importantly, the exponent changes as a function of $\tau$, reflecting the fact that the survival function is not a true power law for a truncated distribution. The effective exponent is shown in Fig. \ref{fig7}(d,e,f) for each simulated case. The effective exponent is $\alpha_e \approx \alpha$ for the unbounded case, indicating the domain size is sufficiently large to approximately represent an infinite power law for this study.

Within the range of $\tau$ where a power law is expected, $\alpha_e$ in Fig. \ref{fig7} becomes increasingly larger than $\alpha$ as $\alpha$ decreases and as the Reynolds number decreases. The decrease in $\alpha$ and Reynolds number both represent an increase in the portion of the $CDF$ that is ``missing'' due to the truncated upper limit of the distribution. For a power law defined only in the range between $\tau_1$ and $\tau_2$, $\alpha_e$ is

\begin{equation}
\alpha_e(\tau) = 1 + \left( \alpha-1 \right) \frac{ \left( \tau/\tau_1 \right)^{1-\alpha} }{ \left( \tau/\tau_1 \right)^{1-\alpha} + \left( \tau_2/\tau_1 \right)^{1-\alpha} }.
\label{eq14}
\end{equation}

\noindent The result $\alpha_e = \alpha$ is recovered when $\tau_2/\tau_1 \rightarrow \infty$, i.e. $Re \rightarrow \infty$. The equation for the truncated distribution in Fig. \ref{fig2} follows a similar form to Eq. \ref{eq14}, except the right-side term in the denominator depends on the prescribed cutoff behavior. In this sense, the lognormal and exponential cutoffs employed here introduce minor quantitative differences in the results, but the scale-dependence of the survival function is a direct consequence of finite-size effects resulting from the finite power law region.

In practice, $\alpha_e$ in Eq. \ref{eq13} can be estimated from discrete histograms approximating the TA interpulse distribution $PDF(\tau)$. For simplicity, a single representative value of $\alpha_e$ is employed here using the average of $\alpha_e(\tau)$ within the inertial subrange where power law statistics are fitted.

\begin{figure}
\centering
\includegraphics{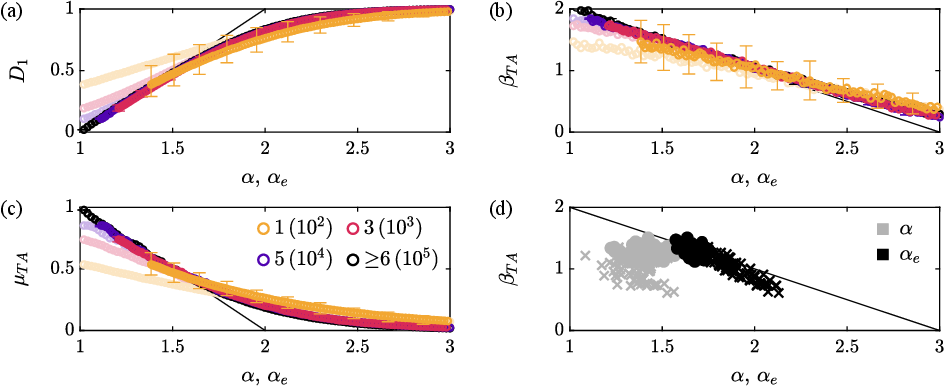}
\caption{Power law exponent as a function of $\alpha$ (transparent markers) and the effective exponent $\alpha_e$ (opaque) for simulated TA signals. (a) Fractal dimension $D_1$. (b) Spectrum exponent $\beta_{TA}$. (c) Intermittency exponent $\mu_{TA}$. (d) Spectrum exponent $\beta_{TA}$ for boundary layer turbulence measurements of the streamwise ($\bullet$) and wall-normal ($\times$) velocity components.}
\label{fig8}
\end{figure}

Fig. \ref{fig8} compares the power law exponents as a function of $\alpha$ and $\alpha_e$. Substituting for the effective exponent $\alpha_e$ accounts for the trends observed in panel (c) of  Figs. \ref{fig3}, \ref{fig4}, and \ref{fig5}. Further, the corrected results follow the linear relations exhibited by the unbounded case. The remaining deviation in the $Re_\lambda \sim 10^2$ case is within the extent of the error bars and may be due to the strong departure from self-similar statistics previously discussed.

As a practical example, Fig. \ref{fig8}(d) shows the correction of $\alpha$ for measured turbulent TA signals. The signals were acquired from a range of positions within a wind tunnel boundary layer above both smooth and rough surfaces \cite{Heisel2020}. The results for the spectrum exponent $\beta_{TA}(\alpha_e)$ align with the expected linear relation and exhibit reduced scatter compared to $\beta_{TA}(\alpha)$. The observed difference in $\beta_{TA}$ between the streamwise and wall-normal velocity components is due to the smaller integral length for the wall-normal component, leading to a narrower inertial subrange and a greater effect of truncation. The effective exponent $\alpha_e$ in Fig. \ref{fig8}(d) successfully accounts for this difference.

The discrepancy between $\beta_{TA}(\alpha)$ in Fig. \ref{fig8}(d) and $\beta_{TA}=3-\alpha$ has been previously explained as an effect of intermittency. Specifically, the correction $\beta_{TA} = 3-\mu_{TA}/2 - \alpha$ was proposed \citep{Bershadskii2004}. The present simulations demonstrate that $\beta_{TA} = 3 - \alpha$ is applicable to intermittent, self-similar processes governed by an unbounded power law. Adjusting the unbounded power law in Fig. \ref{fig4} for intermittency would lead to incorrect results. Rather, the relations between the power law statistics and $PDF(\tau)$ must consider the effective exponent $\alpha_e$ in Eq. \eqref{eq13} resulting from a finite Reynolds number. The success of the correction in Fig. \ref{fig8} demonstrates that the distortion of the survival function propagates to statistics based on the simulated signals, and that the resulting parameters $D_1$, $\beta_{TA}$, and $\mu_{TA}$ depend on the survival function and its effective exponent $1-\alpha_e$.

\section{Summary}
\label{sec5}

For a binary stochastic process described by an unbounded power law probability distribution, linear equations exist to relate the power law exponent of various statistics. The fractal dimension $D_1$, energy spectrum exponent $\beta_{TA}$, and intermittency exponent $\mu_{TA}$ are related as $D_1 = 2-\beta_{TA} = 1-\mu_{TA}$. The relation between $D_1$ and $\mu$ matches the predicted analytical solution. The statistics are also linearly related to the probability exponent $\alpha$, e.g. as $\beta_{TA}=3-\alpha$, for the range of $\alpha$ applicable to turbulent signals ($\alpha \approx$ 1.5). While a small selection of statistics are evaluated here, similar linear relations and trends are expected for other parameters like the correlation integral \cite{Grassberger1983}. 

These relations are directly applicable to the telegraph approximation (TA) of turbulent crossing signals and estimates of the fractal dimension from single-point measurements. However, certain aspects of the results depend on the Reynolds number and the extent of the inertial subrange where power law behavior is expected, in which a finite Reynolds number yields truncated power law statistics. Specifically, a finite Reynolds number changes the effective exponent $\alpha_e$ of the survival function for the TA interpulse $\tau$, which propagates to ensuing statistics. The original linear relations between $\alpha$ and the other exponents can be recovered by considering the effective exponent $\alpha_e$ in Eq. \eqref{eq13}. Yet, cumulative statistics such as the fractal dimension are not self-similar, as $\alpha_e(\tau)$ is strongly scale-dependent when the Reynolds number is small. The departure from self-similarity may be even greater in practice, as the effect of an exponential cutoff on the scale invariance of the power law is not considered here \cite{Laherrere1998}.

These finite Reynolds number effects may help to explain experimental observations in turbulence. Previous findings on the scale dependence of the fractal dimension \cite{Miller1991,Praskovsky1993} are likely due to a combination of the technical challenges discussed in the introduction \cite{Iyer2020} and the finite-size effects studied here. Additionally, deviations from $\beta_{TA}=3-\alpha$ in turbulent TA signals (Fig. \ref{fig8}) are well-described by the proposed correction derived from the survival function. The Reynolds number required for these statistical effects to become negligible, $Re_\lambda \gtrsim 10^5$, is well beyond current numerical and laboratory capacities. These findings are specific to one-dimensional signals, as the extension of the statistical effects to higher dimensions is not explored here.

Based on the design of the simulations, the linear trends and effective exponent discussed above are purely statistical properties of truncated power laws. The findings are independent of the governing Navier-Stokes equations or underlying mechanisms such as self-organized criticality. The relations provide no information on causality, and the only prerequisite is for self-similarity to exist within the signal. In this regard, the conclusions apply to any binary process defined by a truncated power law probability distribution. Importantly, the linear expressions cannot be directly applied to the full turbulent signals exhibiting amplitude variability beyond 0 or 1. In this case, certain relations also depend on the phase of the signal \cite{Higuchi1990}.

\begin{acknowledgments}
The author is financially supported by the US National Science Foundation (NSF-AGS-2031312). The author is also grateful to Professors M. Chamecki and G. G. Katul for fruitful discussions that helped to shape this work. 
\end{acknowledgments}

\bibliography{references}

\end{document}